\documentclass[prb,twocolumn,superscriptaddress,showpacs,epsf,preprintnumbers]{revtex4}

\usepackage{ulem} 
\usepackage{epsfig}
\usepackage{color}

\usepackage{graphicx}
\usepackage{latexsym}
\usepackage{amsmath}
\usepackage{amssymb}
\usepackage{amsfonts}

\newcommand{\gapprox}{{\scriptscriptstyle\stackrel{>}{\sim}}}

\newif\ifgraph

\graphtrue                   %

\begin{document}

\title{Domain-wall and reverse-domain superconducting states of a Pb thin-film bridge \\ on
a ferromagnetic BaFe$_{12}$O$_{19}$ single crystal}

\author{R.~Werner}
\affiliation{Physikalisches Institut -- Experimentalphysik II and
Center for Collective Quantum Phenomena in LISA$^+$,
Universit\"{a}t T\"{u}bingen, Auf der Morgenstelle 14, 72076
T\"{u}bingen, Germany}

\author{A.Yu.~Aladyshkin}
\affiliation{INPAC -- Institute for Nanoscale Physics and
Chemistry, K.U. Leuven, Celestijnenlaan 200D, B--3001 Leuven,
Belgium}
\affiliation{Institute for Physics of Microstructures RAS, 603950,
Nizhny Novgorod, GSP-105, Russia}

\author{S.~Gu\'{e}non}
\affiliation{Physikalisches Institut -- Experimentalphysik II and
Center for Collective Quantum Phenomena in LISA$^+$,
Universit\"{a}t T\"{u}bingen, Auf der Morgenstelle 14, 72076
T\"{u}bingen, Germany}

\author{J.~Fritzsche}
\affiliation{INPAC -- Institute for Nanoscale Physics and
Chemistry, K.U. Leuven, Celestijnenlaan 200D, B--3001 Leuven,
Belgium}

\author{I.M.~Nefedov}
\affiliation{Institute for Physics of Microstructures RAS, 603950,
Nizhny Novgorod, GSP-105, Russia}

\author{V.V.~Moshchalkov}
\affiliation{INPAC -- Institute for Nanoscale Physics and
Chemistry, K.U. Leuven, Celestijnenlaan 200D, B--3001 Leuven,
Belgium}

\author{R.~Kleiner}
\affiliation{Physikalisches Institut -- Experimentalphysik II and
Center for Collective Quantum Phenomena in LISA$^+$,
Universit\"{a}t T\"{u}bingen, Auf der Morgenstelle 14, 72076
T\"{u}bingen, Germany}

\author{D. Koelle}
\email{koelle@uni-tuebingen.de} \affiliation{Physikalisches
Institut -- Experimentalphysik II and Center for Collective
Quantum Phenomena in LISA$^+$, Universit\"{a}t T\"{u}bingen, Auf
der Morgenstelle 14, 72076 T\"{u}bingen, Germany}

\date{\today}
\begin{abstract}
We report on imaging of the nonuniform superconducting states in a
Pb thin film bridge on top of a ferromagnetic BaFe$_{12}$O$_{19}$
single crystal with a single straight domain wall along the center
of the bridge by low-temperature scanning laser microscopy. We
have visualized domain wall superconductivity (DWS) close to the
critical temperature of Pb, when the Pb film above the domain wall
acts as a superconducting path for the current. The evolution of
the DWS signal with temperature and the external-field-driven
transition from DWS to reverse domain superconductivity was
visualized.
\end{abstract}



\maketitle

It is well known that so-called surface or bound states can be
generated by the presence of boundaries in a material. For
example, the formation of surface states for a single electron
wave function in a semi-infinite crystalline lattice due to the
modification of the boundary conditions was described by Tamm
\cite{Tamm32} and by Shockley.\cite{Shockley39} Other examples of
bound states are surface plasmons, propagating along the interface
between a dielectric and a
metal,\cite{Ritchie57,Landau-Lifschitz-EDcm84,Maier07} and surface
acoustic waves traveling along the surface of a material
exhibiting elasticity.\cite{Rayleigh85,Oliner78} In both latter
cases these waves are confined in the direction perpendicular to
the wave vector, i.e.~their amplitudes decay exponentially far
from the interface/surface. The formation of surface bound states
for the superconducting order parameter wave function $\Psi$ was
first considered by Saint-James and de
Gennes.\cite{Saint-James63,Saint-James69} They showed that
localized superconductivity at a superconductor (S)/vacuum or
S/insulator interface can appear at an applied magnetic field
$H_\mathrm{ext}$ above the upper critical field $H_\mathrm{c2}$
for bulk superconductivity. Similarly to this surface
superconductivity, localized superconductivity can also nucleate
near the sample edge in a thin semi-infinite superconducting film
\cite{White66} or in a thin superconducting disk of very large
diameter \cite{Aladyshkin07} in a {\it perpendicular} magnetic
field. Such so-called edge superconductivity (ES), with transition
temperature $T_\mathrm{c}^\mathrm{ES}$, has the same phase
transition line as surface superconductivity,\cite{Tinkham96}
given by $1-T_\mathrm{c}^\mathrm{ES}/T_\mathrm{c0} \simeq 0.59\,
|H_\mathrm{ext}|/H_\mathrm{c2}^{(0)}$. Here, $T_\mathrm{c0}$ is
the superconducting transition temperature in zero magnetic field,
$H_\mathrm{c2}^{(0)}=\Phi_0/(2\pi\xi_0^2)$ and $\xi_0$ are the
upper critical field and coherence length at temperature $T=0$,
respectively, and \mbox{$\Phi_0=\pi\hbar c/e$} is the magnetic
flux quantum. This means that ES will survive up to the critical
field $H_\mathrm{c3}=1.69 H_\mathrm{c2}$, while above
$H_\mathrm{c2}=H_\mathrm{c2}^{(0)} (1-T/T_\mathrm{c0})$ bulk
superconductivity is totally suppressed.

An alternative way to prepare localized states in superconducting
films is to confine the order parameter wave function by a
nonuniform magnetic field in hybrid S/ferromagnet (F) structures
(see e.g.~Ref.~\cite{Aladyshkin09a} and references therein).
Buzdin and Mel'nikov \cite{Buzdin03a} considered a step-like
distribution $b_z(x)=B_0\,{\rm sgn}(x)$ of the perpendicular
component of the magnetic field, $B_z=H_\mathrm{ext}+b_z$, induced
by domain walls in the ferromagnet (with the $z$-axis
perpendicular to the film surface). They demonstrated that
superconductivity will survive in vicinity along the step, even if
the amplitude of the nonuniform magnetic $B_0>H_\mathrm{c2}$. The
dependence of the transition temperature
$T_\mathrm{c}^\mathrm{DWS}(H_\mathrm{ext})$ for {\it domain-wall
superconductivity} (DWS) in a plain superconducting film
(i.e.~infinite in lateral direction) can be estimated as
$1-T_\mathrm{c}^\mathrm{DWS}/T_\mathrm{c0} \simeq \{0.59 -
0.70(H_\mathrm{ext}/B_0)^2 +
0.09(H_\mathrm{ext}/B_0)^4\}B_0/H_\mathrm{c2}^{(0)}$.\cite{Aladyshkin03}

\begin{figure}[tbh]
\includegraphics[width=8.3cm]{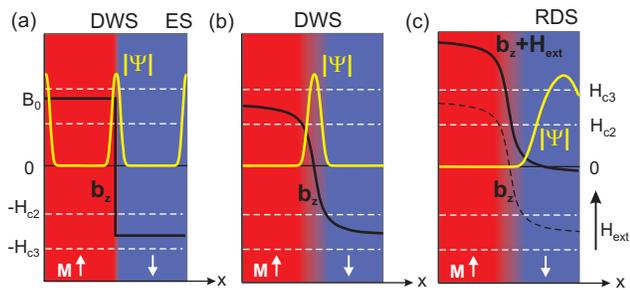}
\caption{(color online). Illustration of DWS, ES and RDS across a
thin film S strip on top of a F substrate with two domains with
perpendicular Magnetization $\bm M$. Magnetic field profiles
$B_\mathrm{z}(x)=H_\mathrm{ext}+b_\mathrm{z}(x)$ inside the S
strip generated by the domains underneath and modulus of
superconducting order parameter $|\Psi(x)|$ are shown for (a)
step-like $b_z(x)$ for $H_\mathrm{ext}=0$, (b) field profile with
finite width for $H_\mathrm{ext}=0$ and (c) finite
$H_\mathrm{ext}\approx B_0$. White dashed lines indicate upper
critical fields $\pm H_\mathrm{c2}$ and $\pm H_\mathrm{c3}$. }
\label{Fig1}
\end{figure}

For flux-coupled S/F structures of finite lateral size the
localized states of ES and DWS may compete as illustrated in
Fig.~\ref{Fig1} for the case of a thin film S strip of finite
width above a F substrate with a domain wall along the center of
the bridge, for $H_{c2}<B_0<H_{c3}$. For a domain structure with
step-like $b_z(x)$ profile and $H_\mathrm{ext}=0$,  ES and DWS
nucleate simultaneously in the S strip as shown in
Fig.~\ref{Fig1}(a). Figure \ref{Fig1}(b) shows the case of a
domain wall with finite width and $H_\mathrm{ext}=0$. Here, DWS
becomes energetically more favorable compared to ES and only DWS
nucleates.\cite{Comment:GLmodel} For $H_\mathrm{ext}\neq 0$, the
local field is compensated above the domain with magnetization
direction opposite to $H_\mathrm{ext}$. If $||H_\mathrm{ext}|-B_0|
<H_\mathrm{c2}$ superconductivity is turned on above this reverse
domain while it is still suppressed above the parallel domain
[c.f.~Fig.~\ref{Fig1}(c)]. This effect is termed {\it
reverse-domain superconductivity} (RDS)
.\cite{Yang04b,Fritzsche06} We note that when $H_\mathrm{c2}$
becomes larger than $|H_\mathrm{ext}|+B_0$ (e.g.~upon cooling)
above the parallel domain, superconductivity may also nucleate
there and the entire strip will be in the superconducting state,
which we call {\it complete superconductivity} (CS).

First fingerprints of RDS and DWS have been found by electric
transport measurements on S/F hybrids with a rather complex domain
structure in BaFe$_{12}$O$_{19}$ (BFO) crystals \cite{Yang04b} and
multilayered CoPt films \cite{Aladyshkin10a} with perpendicular
magnetic anisotropy. Using low-temperature scanning laser
microscopy (LTSLM), RDS has been visualized in a hybrid
Nb/PbFe$_{12}$O$_{19}$ system.\cite{Fritzsche06} However, due to
the complex domain structure and relatively small domain size,
visualization of DWS was not possible. Recently, significant
improvements have been achieved, regarding the fabrication of
specially polished BFO crystals, characterized by a well defined
and stable domain structure with straight domain walls separated
by typically $30\,\mu$m.\cite{Fritzsche09a,Aladyshkin09} Here we
report on the direct imaging of the development of DWS and RDS in
a hybrid S/F structure, consisting of a superconducting Pb film on
top of a ferromagnetic BFO crystal by means of
LTSLM.\cite{Zhuravel06,Lukashenko06,Wagenknecht06,Fritzsche06,Wang09}

\begin{figure*}[tbh]
\includegraphics[width=16.8cm]{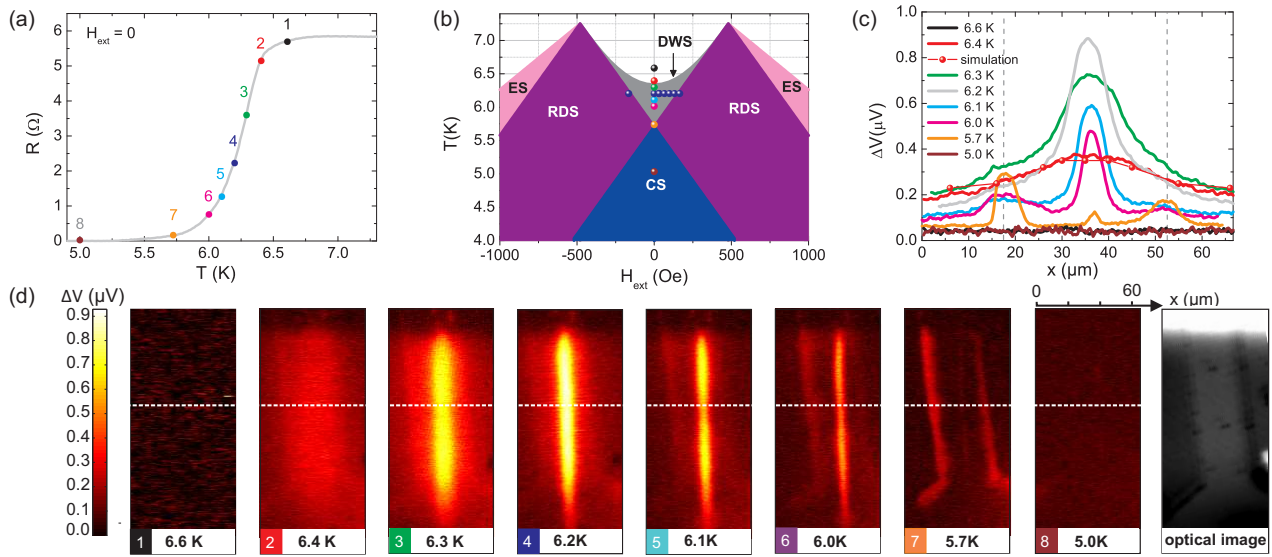}
\caption{Evolution of DWS upon cooling through $T_\mathrm{c}$ and
$H_\mathrm{ext}-T$ phase diagram. (a) $R(T)$ curve
($I=100\,\mu$A); dots indicate bias points of LTSLM voltage images
1--8 in (d) and corresponding line scans in (c). (b),
${H_\mathrm{ext}-T}$ phase diagram, constructed from
experimentally determined values $T_\mathrm{c0}=7.25\,$K,
$B_0=480\,$G and $H_\mathrm{c2}^{(0)}=2.25\,$kOe. The phase
diagram contains separate regions of DWS, ES, RDS and CS. Dots
label bias points for LTSLM data shown in (c), (d) and
Fig.~\ref{Fig3}. (c), line scans $\Delta V(x)$ across the bridge
for different $T$, taken from voltage images in (d). Red dots show
simulation results for $T=6.4\,$K. The position of the edges of
the bridge is indicated by dashed grey lines. (d), series of LTSLM
voltage images $\Delta V(x,y)$ (1--8 from left to right) taken at
different$T$ during cooling the Pb bridge through its resistive
transition ($I=10\,\mu$A). White dashed lines indicate the
position of line scans in (c). The graph on the right shows a
corresponding optical LTSLM image. \label{Fig2}}
\end{figure*}

We prepared a $40\,$nm thick and 30~$\mu$m wide Pb microbridge on
top of a BFO substrate, so that only a single domain wall is
running along the center of the Pb bridge parallel to the current
flow. The BFO substrate and the Pb thin film were separated by a
4\,nm thick insulating Ge layer so that the system is only
flux-coupled. From  resistance $R$ vs $H_\mathrm{ext}$
measurements at variable $T$ of a reference sample with several
domain walls oriented perpendicular to the long side of the
bridge,\cite{Aladyshkin-crossover-SF} we compose the
$H_\mathrm{ext}-T$ phase diagram shown in Fig.~\ref{Fig2}(b).

For imaging by LTSLM, the sample was mounted on the cold finger of
a Helium gas flow cryostat, with an optical window to enable
irradiation of the sample surface in the $(x,y)$ plane by a
focused laser beam with beam spot diameter $\sim
1.5-2\,\mu$m.\cite{Wagenknecht06,Wang09} The amplitude modulated
laser beam (at frequency $f\approx 10\,$kHz) induces a local
increase of temperature $\delta T(x-x_0,y-y_0)$ centered at the
beam spot position $(x_0,y_0)$ on the sample surface. During
imaging, the Pb bridge is biased at a constant current $I$, and
the beam-induced change of voltage $\Delta V(x_0,y_0)$ is recorded
with lock-in technique. The LTSLM voltage signal can be
interpreted as follows: If the irradiated part of the sample was
in the normal state with resistivity $\rho_n$, the laser beam
induces a very small voltage signal $\Delta V\propto\partial
\rho_n /\partial T$. However, if the irradiated region took part
in the transfer of a substantial part of the supercurrents, the
beam-induced suppression of superconductivity might switch the
sample from a low-resistive state to a high-resistive state. This
effect should be maximal if $I$ is close to the overall critical
current $I_c=I_c(T,H_\mathrm{ext})$ of the sample. In this case
LTSLM allows one to map out the ability of the sample to carry
supercurrents.

In order to trace out the evolution of DWS with temperature, we
recorded a series of LTSLM voltage images $\Delta V(x,y)$ at
$H_\mathrm{ext}=0$ and different $T$ across the resistive
transition of the Pb bridge.

Figure~\ref{Fig2}(a) shows the $R(T)$ curve of the Pb/BFO
microbridge; the labels 1--8 indicate the bias points for which
LTSLM images and line scans are shown in Figs.~\ref{Fig2}(d) and
\ref{Fig2}(c), respectively. The dots in the $H_\mathrm{ext}-T$
phase diagram shown in Fig.~\ref{Fig2}(b) indicate the bias points
for which LTSLM data are shown. The LTSLM voltage images 1--8 in
Fig.~\ref{Fig2}(d) show the evolution of the superconducting
properties of the Pb/BFO bridge upon cooling through
$T_\mathrm{c}$ (from left to right) at $H_\mathrm{ext}$=0;
according to Fig.~\ref{Fig2}(b), these should cover the
transitions from the normal state to DWS and finally to CS. The
graph on the right shows an optical LTSLM image, in order to
indicate size and position of the bridge in the voltage
images.\cite{Comment:Optical} For a more quantitative analysis, in
Fig.~\ref{Fig2}(c) we show line scans $\Delta V(x)$ across the
bridge [along the white dashed lines in Fig.~\ref{Fig2}(d)].

Starting with the highest temperature $T=6.6\,$K, the voltage
image in Fig.~\ref{Fig2}(d) and the corresponding linescan (black
line) in Fig.~\ref{Fig2}(c) shows no signal, as the bridge is in
the normal state. Lowering $T$ to 6.4\,K [entering the resistive
transition shown in Fig.~\ref{Fig2}(a)], the voltage image gives a
small homogeneous signal with a broad maximum centered above the
bridge [red line in Fig.~\ref{Fig2}(c)]. For a (still) resistive
Pb bridge with homogeneous conductivity but finite $\partial
R/\partial T$, this behavior can be simply explained by the finite
width of the beam-induced $\delta T(x,y)$ profile, i.e.~its tails
will induce a voltage signal, even if the beam spot is positioned
outside the bridge. This is confirmed by numerical simulations
[c.f.~red data points in Fig.~\ref{Fig2}(c)], which solve the heat
diffusion equation for an absorbed laser power of $25\,\mu$W, a
beam spot diameter of $2\,\mu$m and thermal conductivity of the
BFO substrate of $\rm 0.8\,Wcm^{-1}K^{-1}$. These simulations
yield a maximum increase in beam-induced temperature $\Delta T=
0.14\,$K.

Upon further cooling (see voltage images and corresponding line
scans for $T=6.3\,K$ and $T=6.2\,K$), a clear LTSLM signal
develops, running along the domain wall [green and blue lines,
respectively, in Fig.~\ref{Fig2}(c)]. This observation can be
interpreted as an evidence that a channel above the domain wall
with  higher conductivity than the regions above the domains has
formed, and therefore the current density $j(x)$ has a maximum
above the domain wall. We note that, although according to the
$H_\mathrm{ext}-T$ phase diagram the sample should be in the the
DWS state, the overall resistance of the bridge is close to the
full normal resistance. This is consistent with numerical
simulations based on the time-dependent Ginzburg-Landau equations,
which indicate that for our experimental situation the critical
current density $j_\mathrm{c,DWS}$ along the domain-wall channel
is too small, i.e.~the bias current might be above the critical
current of this channel. This also explains why, upon decreasing
$T$, the LTSLM signal at the domain wall increases, as
$j_\mathrm{c,DWS}$ increases, and the peak in $\Delta V(x)$
becomes sharper (see below). We did not find a similar enhancement
of the LTSLM voltage signal at the edges of the bridge, i.e.~we do
not find any signature of ES. We attribute this to the finite
width of the domain wall, which stabilizes DWS compared to ES.

For $T$ $<$ 6.2\,K the amplitude of the peak of the LTSLM response
at the domain wall decreases as $T$ decreases, and the maximum of
the LTSLM signal shifts towards the edges of the bridge; see
magenta and orange lines in Fig.~\ref{Fig2}(c) for $T=6.0\,K$ and
$T=5.7\,K$, respectively, and the corresponding voltage images in
Fig.~\ref{Fig2}(d). We interpret this observation as the
transition from DWS to CS, which is consistent with the phase
diagram shown in Fig.~\ref{Fig2}(b). At this transition, CS
spreading over the whole sample becomes favorable and the sample
is turned into the mixed state. The onset of CS can explain the
appearance of two pronounced maxima in $\Delta V(x)$ at the sample
edges: In the mixed state the current distribution depends on the
edge energy barrier for vortex entry. Upon laser irradiation, the
edge energy barrier is locally suppressed, which in turn opens a
gate for vortex entry/exit. Hence one can expect that irradiation
at the edges of the bridge should strongly affect the vortex
pattern and the resulting current distribution. In contrast, laser
irradiation of the interior of the bridge does not change the
existing energy barrier, and the modification of the current
pattern is probably less pronounced, and therefore the
beam-induced voltage change is much smaller. Finally, at
$T=$5.0\,K the LTSLM signal is zero [c.f.~Fig.~\ref{Fig2}(d) and
brown line in Fig.~\ref{Fig2}(c)], which indicates that the bridge
is in the CS state and the beam-induced perturbation is not strong
enough to suppress superconductivity and to induce a voltage
signal.

\begin{figure}[b]
\includegraphics[width=8.3cm]{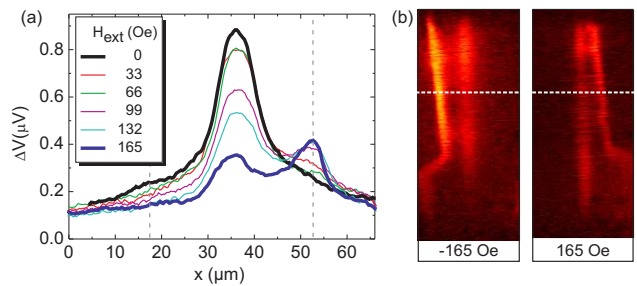}
\caption{(color online). Switching from DWS to RDS: variable
$H_\mathrm{ext}$ at  $T=6.2\,$K. (a) Line scans $\Delta V(x)$
along white dashed line in (b) for different $H_\mathrm{ext}\ge0$;
dashed grey lines indicate edges of the bridge. (b) LTSLM images
for maximal $|H_\mathrm{{ext}}|$. } \label{Fig3}
\end{figure}

Finally, we investigated the effect of finite perpendicular field
$|H_\mathrm{ext}|\le 165\,$Oe on superconductivity in our system.
The measurements were carried out at $T=6.2$\,K, which corresponds
to the most pronounced LTSLM signal above the domain wall at
$H_\mathrm{ext}=0$. Figure \ref{Fig3}(a) shows the evolution of
the LTSLM voltage signal $\Delta V(x)$ with increasing external
field for positive polarity. For $H_\mathrm{ext}=0$ the DWS signal
is clearly visible as described above. With increasing
$H_\mathrm{ext}$ the amplitude of the domain-wall signal decreases
monotonously while its width stays roughly constant.
Simultaneously a signal above the reverse (right) domain appears.
In the RDS state, for $H_\mathrm{ext}\gapprox 70\,$Oe, the voltage
signal shows a peak at the right edge of the bridge, which can be
explained in the same way as for the edge signal discussed in the
context of the $T$-series shown in Fig.~\ref{Fig2}. Figure
\ref{Fig3}(b) shows LTSLM voltage images taken at
$H_\mathrm{ext}=-165\,$Oe (left image) and
$H_\mathrm{ext}=+165\,$Oe (right image), which clearly demonstrate
switching between the RDS states above the two domains upon
reversing the external field polarity.

In conclusion, we have clearly identified the formation of the
spatially inhomogeneous superconducting state in a superconducting
Pb thin film induced by the stray field of the domains in the
ferromagnetic substrate BFO underneath. The crucial feature of the
investigated system is that the superconducting Pb bridge was
fabricated exactly above a single straight domain wall, which is
running along the center of the bridge. Such a well-defined
geometry of the hybrid Pb/BFO sample makes it possible to directly
visualize the localized and delocalized superconductivity by means
of low-temperature scanning laser microscopy. We imaged the
evolution of DWS with decreasing temperature. Using the external
field as a control parameter, we demonstrated that
superconductivity in a wide superconducting bridge can be switched
from the DWS to RDS state. This opens up interesting perspectives
for the creation of spatially nonuniform superconducting states
and for their manipulation by external and ''internal`` magnetic
fields.

This work was funded by the DFG via Grant No.~KO 1303/8-1, the
Methusalem Funding of the Flemish Government, the NES -- ESF
program, the Belgian IAP, the Fund for Scientific Research --
Flanders (F.W.O.--Vlaanderen), the RFBR, RAS under the Program
``Quantum physics of condensed matter", and FTP ``Scientific and
educational personnel of innovative Russia in 2009--2013". R. W.
gratefully acknowledges support by the Cusanuswerk, Bisch\"ofliche
Studienf\"orderung.


\begin{references}
\bibitem{Comment:GLmodel}
We calculate $|\Psi(x)|$ by solving numerically the linearized
Ginzburg-Landau Eq.~for a thin superconducting bridge of finite
width in the presence of the nonuniform field
$B_z(x)=H_\mathrm{ext}+(2B_0/\pi)\,\arctan(x/L)$, following the
approach of Ref.~\cite{Aladyshkin03}. The case $L\to 0$
corresponds to a step-like field profile, and any (even very small
but) finite $L$ gives a finite transient region for the stray
magnetic field near the domain wall. I.e. for the system
investigated here, with typical width 5-10\,nm of the domain walls
in the BaFe$_{12}$O$_{19}$ substrate, ES will be suppressed.
\bibitem{Comment:Optical}
The bright area in the upper part of the optical image is due to
an Au contact layer on top of the Pb film, which gives a strongly
enhanced reflected laser signal and due to reduced absorbtion a
reduction of the LTSLM signal which can be clearly seen in the
images 2--7 in Fig.~\ref{Fig2}(d).
\end{references}
\end{document}